\documentclass{aa}  

\usepackage{graphicx}

\usepackage{txfonts}
\usepackage{CJK}
\usepackage{hyperref}

\begin{document}
\begin{CJK*}{UTF8}{bsmi}

    \title{Interaction of the central jet with the surrounding
    gas in the protostellar outflow from IRAS 04166+2706}

   \titlerunning{Momentum transfer from the jet in the IRAS 04166 outflow}

   \author{M. Tafalla\inst{1} \and D. Johnstone\inst{2,3} \and 
   J. Santiago-Garc\'{\i}a\inst{4} \and Q. Zhang\inst{5} \and
   H. Shang (尚賢)\inst{6}
   \and C.-F. Lee \inst{6}
   }

    \institute{Observatorio Astron\'omico Nacional (IGN), Alfonso XII 3,
       E-28014 Madrid, Spain \\
    \email{m.tafalla@oan.es}
    \and
    NRC Herzberg Astronomy and Astrophysics, 5071 West Saanich Rd, Victoria, BC, V9E 2E7, Canada
    \and 
    Department of Physics and Astronomy, University of Victoria, Victoria, BC, V8P 5C2, Canada
    \and
    Instituto de Radioastronom\'{\i}a  Milim\'etrica (IRAM), Av. Divina Pastora 7, N\'ucleo Central, E-18012 Granada, Spain
    \and
    Center for Astrophysics, Harvard \& Smithsonian, 60 Garden Street, Cambridge, MA 02138, USA
    \and
    Institute of Astronomy and Astrophysics, Academia Sinica, Taipei 106319, Taiwan
    }

   \date{Received ; accepted}

  \abstract
{The outflow from the Class 0 protostar IRAS 04166+2706 (hereafter IRAS 04166) 
   contains a remarkably symmetric jet-like component of extremely  high-velocity (EHV) gas.}
   {We studied the IRAS 04166 
   outflow and investigated the relation between its EHV component 
   and the slower outflow gas.}
   {We mosaicked the CO(2--1) emission from the IRAS 04166 outflow
   using the 12m and the Compact Arrays of ALMA. 
   We also developed a ballistic toy model of the
   gas ejected laterally from a jet to interpret the data.}
   {In agreement with previous observations, the ALMA data show that
   the slow outflow component is distributed in two opposed conical lobes
   and has a shear-flow pattern with velocity increasing toward the axis.
   The EHV gas consists of a series of arc-like condensations that span the full width
   of the conical lobes and merge with their walls, suggesting
   that the fast and slow outflow components are physically connected.
   In addition, position--velocity diagrams along
   the outflow axis show finger-like extensions that connect
   the EHV emission with the origin of the diagram,
   as if part of the EHV gas had been decelerated by 
   its interaction with the low-velocity outflow.
   A ballistic model can reproduce these finger-like extensions
   assuming that the EHV gas consists of
   jet material that has been ejected laterally 
   over a short period of time and has transferred part of its 
   momentum to the surrounding shear flow. }
   {The EHV gas in the IRAS 04166 outflow seems to play a role in the acceleration 
   of the slower gas component. 
   The presence of similar finger-like extensions in the position--velocity
   diagrams of other outflows suggests that this process may be 
   occurring in other systems, even if the EHV component is
   not seen because it has an atomic composition.}


   \maketitle
%

\section{Introduction}

Bipolar outflows are ubiquitous in star-forming regions.
They are found in protostars of all masses and environments, 
and seem to represent a necessary 
ingredient in the process of star formation
(see \citealt{fra14}, \citealt{bal16} for 
dedicated reviews).
 
Although they are common and despite multiple decades of research,
important aspects of
bipolar-outflow physics remain poorly understood.
Outflows are thought to be powered by the combined action of
gravity,
disk rotation, and magnetic fields, but 
whether the solution to this complex problem
results in a wide-angle wind 
or a highly collimated jet 
(or some combination of the two) is strongly
debated \citep{rag93a,rag93b,mas93,shu94,sha06,rab22,sha23}.

Outflows powered 
by the youngest protostars often present the highest degree of 
collimation and the highest velocities, and are thought 
to best reflect the intrinsic properties of the underlying 
driving agent \citep{lee20}. 
A subset of these 
young outflows stands out for the presence of a
distinct velocity component that appears 
in the spectra as a detached secondary peak  and is
commonly referred to as the extremely high-velocity 
(EHV) component  (e.g., \citealt{bac96}).
This component presents the highest degree of collimation
and often carries
a significant fraction of the total outflow energy and momentum,
which suggests that it could play a role
in the acceleration of the outflow material \citep{bac90,taf04}.
The EHV component, in addition, presents an unusual
chemical composition rich in SiO and other oxygen-bearing
species \citep{taf10,kri11,tyc21} indicative of a different
origin from the rest of the outflow, which is likely dominated 
by swept-up ambient gas. One possibility 
is that the EHV gas represents a wind from the
protostar or the inner disk since its composition
resembles that predicted by chemical models
\citep{gla91,tab20}.
Further work is clearly needed to understand the physics
and chemistry of the EHV component, and to assess its role in the
larger outflow phenomenon.

The outflow from the Class 0 source IRAS 04166+2706 in Taurus 
(hereafter IRAS 04166), initially discovered by \cite{bon96},
represents an excellent target to investigate the origin and 
nature of the EHV component.
It is at an estimated distance of 156~pc \citep{kro21},
and its central source has a luminosity of 0.4~L$_\odot$
and a mass in the range 0.15--0.39~M$_\odot$
\citep{oha23,phu25}.
No Herbig-Haro objects are associated with it, 
although \cite{dav10} found weak H$_2$ emission along its lobes.
Its EHV component was first identified by \cite{taf04} 
using the IRAM 30m telescope, and was later 
studied by \cite{san09} and \cite{wan14}
using data from the 
Plateau de Bure Interferometer and the Smithsonian Submillimeter Array.
These observations revealed 
that the EHV component of IRAS 04166 consists 
of a jet-like chain of condensations
that are located symmetrically with respect to the 
protostar, as if they  resulted from a series of episodic ejections.
The EHV component, in addition, was found to present a systematic 
sawtooth velocity pattern in position--velocity (PV) diagrams
that resembles those predicted by
models of pulsating jets (\citealt{sto93}, but see 
\citealt{wan19} for an alternative interpretation in terms of a wide-angle wind).

Further observations of two selected EHV condensations in
the IRAS 04166 outflow were carried out by \cite{taf17} using 
the Atacama Large Millimeter/submillimeter Array (ALMA).
Analyzing CO(2--1) data, these authors showed that the EHV gas
is distributed in parabolic structures reminiscent of
bow shocks, and that the material in them 
is expanding radially with a
velocity that increases linearly with distance from 
the outflow axis. This geometry and kinematics
supports the interpretation that the EHV emission 
represents material that has been 
ejected laterally from the jet 
in a series of internal shocks.

To further investigate the
EHV component of the IRAS 04166 outflow, 
we  carried out new ALMA observations to map
the full length of the main flow and cover multiple EHV peaks. 
In this paper we present the results of these observations
and describe how they provide evidence for physical
interaction between the
EHV regime and the slower outflow component.

\section{Observations}

The IRAS 04166 outflow was
observed with ALMA using its Band 6 receiver \citep{edi04,ker04}
between October 2021 and January 2022.
The main goal of the project (reference 2021.1.00575.S)
was to map the CO(2--1) emission from the
outflow, so this line was observed with a bandwidth of 325~km~s$^{-1}$ 
and a velocity resolution of 0.16~km~s$^{-1}$.
Additional correlator windows were used to observe the emission of SiO(5--4),
$^{13}$CO(2--1), and SO(6,5--5,4) with similar velocity
resolutions, and a 2~GHz-wide window was added to observe the 
continuum emission at a frequency of 232~GHz. 
Both the ALMA Compact Array (ACA) and the 12m array 
were used to cover a range of baselines from 8.9 to 976.6 m.
To map the full extent of the emission, mosaics
of 17 and 14 pointings were made with the ACA and 12m arrays, respectively.

The visibilities obtained with the interferometer
were calibrated using the CASA software \citep{cas22} version 6.2.1-7 
and the facility-provided pipeline. 
After calibration, the continuum ACA and 12m array visibilities 
were combined and Fourier transformed using the multi frequency 
synthesis mode and natural weighting for maximum sensitivity.
The resulting image was cleaned using 
the {\tt tclean} command with the
{\tt hogbom} \citep{hog74}
option and no special mask because
the emission was only detected toward the protostar.
The beam size
was $0\farcs 7 \times 0\farcs 6$, and the image rms noise level was
about 60~$\mu$Jy~beam$^{-1}$.

For the spectral-line emission, the
calibrated ACA and 12m-array visibilities were resampled to a velocity resolution 
of 0.5~km~s$^{-1}$, 
continuum-subtracted using line-free channels, and Fourier-transformed.
The resulting maps had a synthesized beam of
$0\farcs 7 \times 0\farcs 6$ and a
typical rms level of 4.4 mJy~beam$^{-1}$
per 0.5~km~s$^{-1}$ channel, equivalent to 0.25~K
in brightness temperature.
The maps were cleaned with {\tt tclean}
using the {\tt multiscale} \citep{cor08}
option (scale sizes of 0, 5, 10, and 20 pixels) and 
{\tt auto-multithresh} masking \citep{kep20}.
Several tests were carried out to optimize the choice of 
imaging parameters,
although the use of 
alternative values for the weighting scheme or the cleaning
options did not produce significantly different maps. 

Once the images were cleaned using CASA, the data were converted to the
GILDAS\footnote{\url{http://www.iram.fr/IRAMFR/GILDAS}} format for
further processing. To test the
ability of the ALMA observations to recover the outflow emission, 
we synthesized a series of
CO(2--1) spectra assuming an $11''$ angular resolution and compared them 
with the IRAM 30m single-dish observations of \cite{taf04}. 
This comparison is presented in the left panel of
\ref{fig_convol}, and shows
that the ALMA observations fully recovered the emission of the EHV outflow regime,
which is the main focus of this study. The ALMA observations, however,
miss significant 
emission at velocities lower than about 10~km~s$^{-1}$ with respect to the
ambient cloud despite including ACA data, suggesting that the lowest-velocity 
outflow emission is very extended.
As we show, this lowest-velocity extended
emission is not critical for the analysis presented here, and 
if necessary, it can be studied using the lower 
angular resolution observations of \cite{san09}.

\section{Results}

\subsection{Continuum data}

\begin{figure*}
        \centering
    \includegraphics[width=\hsize]{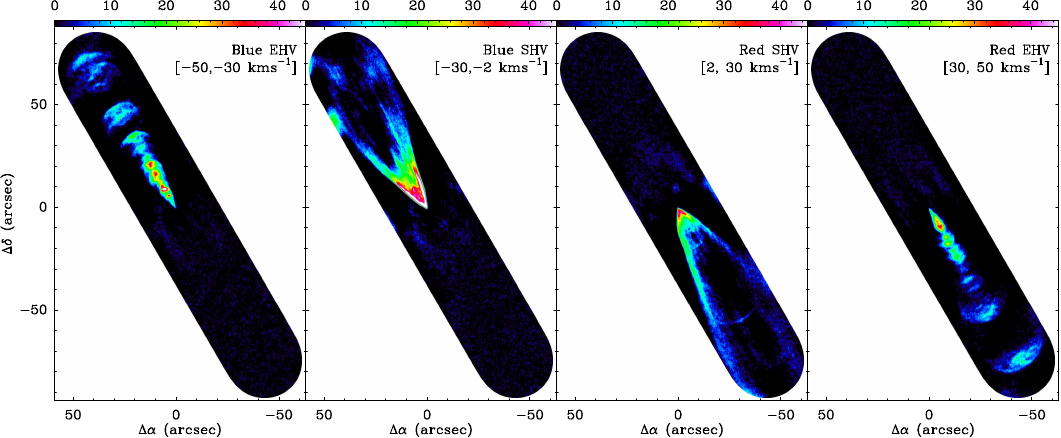}
    \caption{Maps of the CO(2--1) intensity integrated over the EHV and SHV 
    velocity regimes. The velocity ranges of integration are given    
    inside square brackets and are
    measured with respect to a cloud LSR velocity of 
    6.7~km~s$^{-1}$. The map coordinates are offsets measured with 
    respect to the position
    of IRAS 04166, and the colorintensity scales at the top are in units of K~km~s$^{-1}$. 
    We note the different geometry of the SHV and EHV 
    components.
    }
        \label{fig_flow_4v}
\end{figure*}

The map of the 232 GHz continuum emission only 
shows a single point-like source toward the position of the protostar,
and for completeness is shown in Fig.~\ref{fig_cont}.
While no other emission sources are detected, the map
shows a noticeable increase in the level of residuals toward the vicinity of the 
protostar, which likely results from the interferometer filtering out
the extended emission from the dense core 
that surrounds this Class 0 source. The core has 
a diameter of about one arcminute, and has been 
previously mapped at millimeter wavelengths by multiple authors
\citep{mot01,taf04,hac13,bra17,esw21}. 

To characterize the 232~GHz emission peak we 
fit it with a 2D Gaussian using the CASA command {\tt imfit}.
The estimated peak flux density is $59.68\pm 0.06$~mJy~beam$^{-1}$ at the position 
RA(ICRS)= 04$^{\mathrm h}$19$^{\mathrm m}$42\fs505, 
DEC(ICRS) = 27$^\circ$13$'$35\farcs 87, and
the integrated flux is $76.2 \pm 0.1$~mJy. These values are in good agreement with those derived at approximately the same frequency
by \cite{san09} and also by the eDisk ALMA project.
These latter observations have an angular resolution ten times 
higher than ours and reveal a smooth dust disk of approximately 20~au
in radius and an inclination of 47$^\circ$ with respect to the 
line of sight \citep{oha23,phu25}. Even at this highest
angular resolution, the central source seems to be a single object. 
Assuming a typical dust temperature of 20~K, optically thin emission,
an emissivity per dust mass of 0.9~cm$^2$~g$^{-1}$ \citep{oss94}, 
and a standard gas to dust ratio of 100 \citep{boh78},
the estimated flux of 76~mJy corresponds to a disk
mass of 0.03~M$_\odot$. 

\subsection{Overview of the line data}
\label{sect_overview}

As shown by previous observations, the spectra from the IRAS 04166 outflow 
reveal the presence of two distinct velocity regimes
(\citealt{taf04}, \citealt{san09}, \cite{wan14},
and also \ref{fig_convol}).
At low speeds, the spectra present 
a standard outflow wing, so this regime is 
referred to as the standard high-velocity (SHV) gas. At 
higher speeds some spectra present a distinct secondary 
peak that corresponds to the EHV component.
Following previous work on this source,
we define the boundary between the SHV and EHV 
regimes at 30~km~s$^{-1}$
with respect to the ambient cloud ($V_\mathrm{LSR} = 6.7$~km~s$^{-1}$),
and we note that given the continuity that we find between the two 
regimes in both spectra and channel maps, we do not find 
necessary to define an intermediate velocity regime.

To compare the spatial distribution of the SHV and EHV outflow regimes, we present in 
Fig.~\ref{fig_flow_4v} maps of the blue- and redshifted
CO(2--1) emission integrated over the SHV regime (middle panels) and the
EHV regime (outermost panels). 
Overall, the maps show that the outflow
emission has a high degree of symmetry and a strong bipolarity.
They also show that the SHV and EHV regimes have very different geometries.
The SHV emission occupies the inside of two opposed 
conical lobes that have the IRAS source at their vertex and present
similar opening angles.
The EVH emission, in contrast, appears as a
collection of discrete condensations that lie along the axis of the outflow, 
widen with distance from the protostar, and are  bounded by 
the walls of the SHV conical emission.

\begin{figure*}
  \centering
   \includegraphics[width=\hsize]{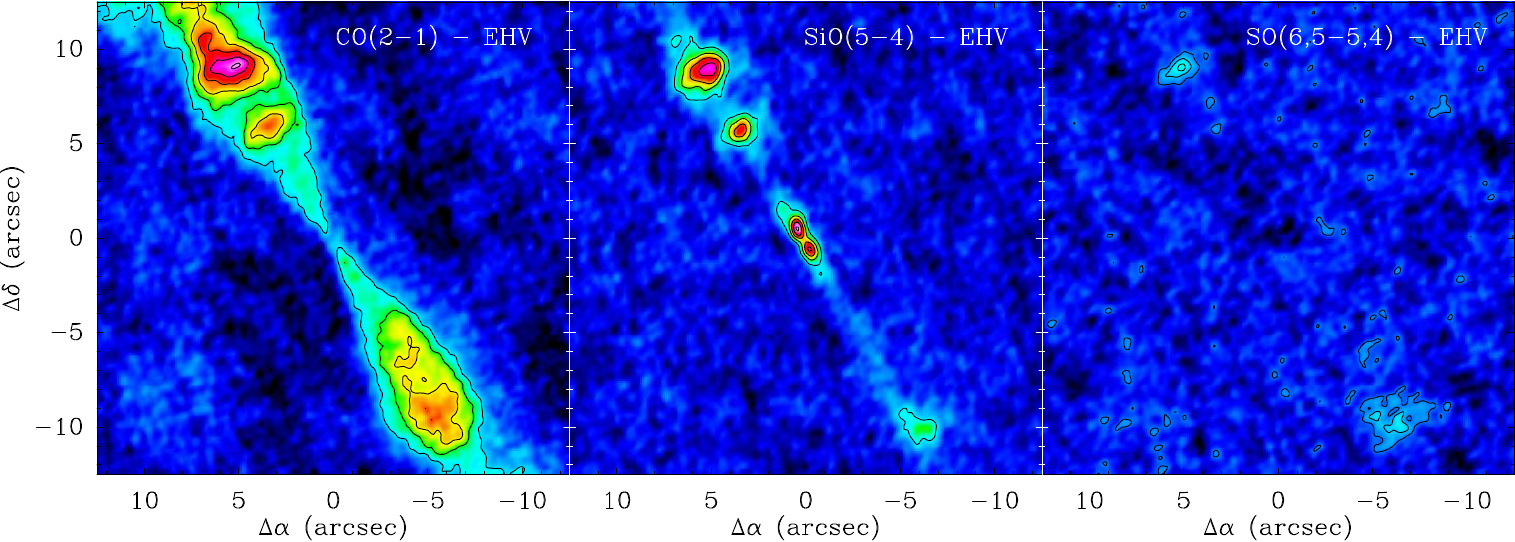}
        \caption{Maps of the combined blueshifted and redshifted EHV emission
        of CO(2--1), SiO(5--4), and SO(6,5--5,4) toward the inner
        $25''\times 25''$ of the IRAS 04166 outflow.
        For CO(2--1) and SiO(5--4), the first contour and contour interval are 9~K~km~s$^{-1}$, while for the weaker SO(6,5--5,4) they are 3~K~km~s$^{-1}$.
        The map offsets are referred to the position of IRAS 04166.}
        \label{fig_ehv_center}
\end{figure*}

Before discussing in more detail the kinematics of the CO(2--1) emission, we 
briefly review the emission of the additional
lines covered by the ALMA observations: SiO(5--4), SO(6,5--5,4), and
$^{13}$CO(2--1). These lines
are significantly weaker than CO(2--1), as can be 
seen in the comparison between spectra presented in Fig.~\ref{fig_convol}.
Since the ALMA integration time was optimized for
CO(2--1) mapping, the emission from the additional lines lacks
sensitivity and therefore 
provides only limited information about the outflow.
This can be seen in Fig.~\ref{fig_flow_others}, which presents
integrated intensity maps of SiO(5--4), SO(6,5--5,4), and 
$^{13}$CO(2--1)
using the same velocity ranges used for CO(2--1) in Fig.~\ref{fig_flow_4v}. 
As the figure shows, none of the additional lines is detected in all four
outflow regimes, and most of the detected emission is concentrated toward 
the vicinity of the central protostar. 

The brightest of our additional lines is SiO(5--4), which traces predominantly 
the EHV regime, as has been previously seen
in outflows with EHV component
\citep{gui92,hir06,cod07,lee07,san09,tyc21}. This preference of the SiO emission
for the EHV component likely reflects the 
peculiar chemical composition of the EHV gas \citep{taf10,tyc21},
and makes the emission of this tracer resemble the EHV component of CO(2--1).
A detailed comparison between the EHV emission of SiO(5--4) and CO(2--1)
reveals significant differences only toward the 
vicinity of the driving source.
This is illustrated in Fig.~\ref{fig_ehv_center}, which presents maps of the 
combined (red+blue) EHV emission of CO(2--1), SiO(5--4), and SO(6,5--5,4)
toward the inner $25''\times 25''$ of the outflow.
As can be seen, the CO(2--1) emission from the protostar vicinity 
is relatively featureless, and is distributed in two conical lobes that end in
condensations resembling bow shocks.
The SiO(5--4) emission, in contrast, presents a jet-like distribution with
bright peaks at each side of the protostar.
As discussed in Sect.~\ref{sect_consequences} this highly collimated
SiO(5--4) emission may be tracing the unperturbed jet component that 
undergoes internal shocks at further distance, although higher angular 
resolution observations are required to test this interpretation.

In contrast with CO(2--1) and SiO(5--4), the SO(6,5--5,4) emission is barely detected
toward the brightest EHV peaks, and no evidence for EHV emission is
seen in $^{13}$CO(2--1). While these tracers may provide additional clues
on the chemical composition of the IRAS 04166 outflow, they add little
information on the outflow kinematics, which is the focus
of this paper. For this reason, from now on we  
focus on the brighter CO(2--1).

\subsection{Velocity structure of the IRAS 04166 outflow}
\label{sec_vel_str}

\begin{figure*}
        \centering
    \includegraphics[width=\hsize]{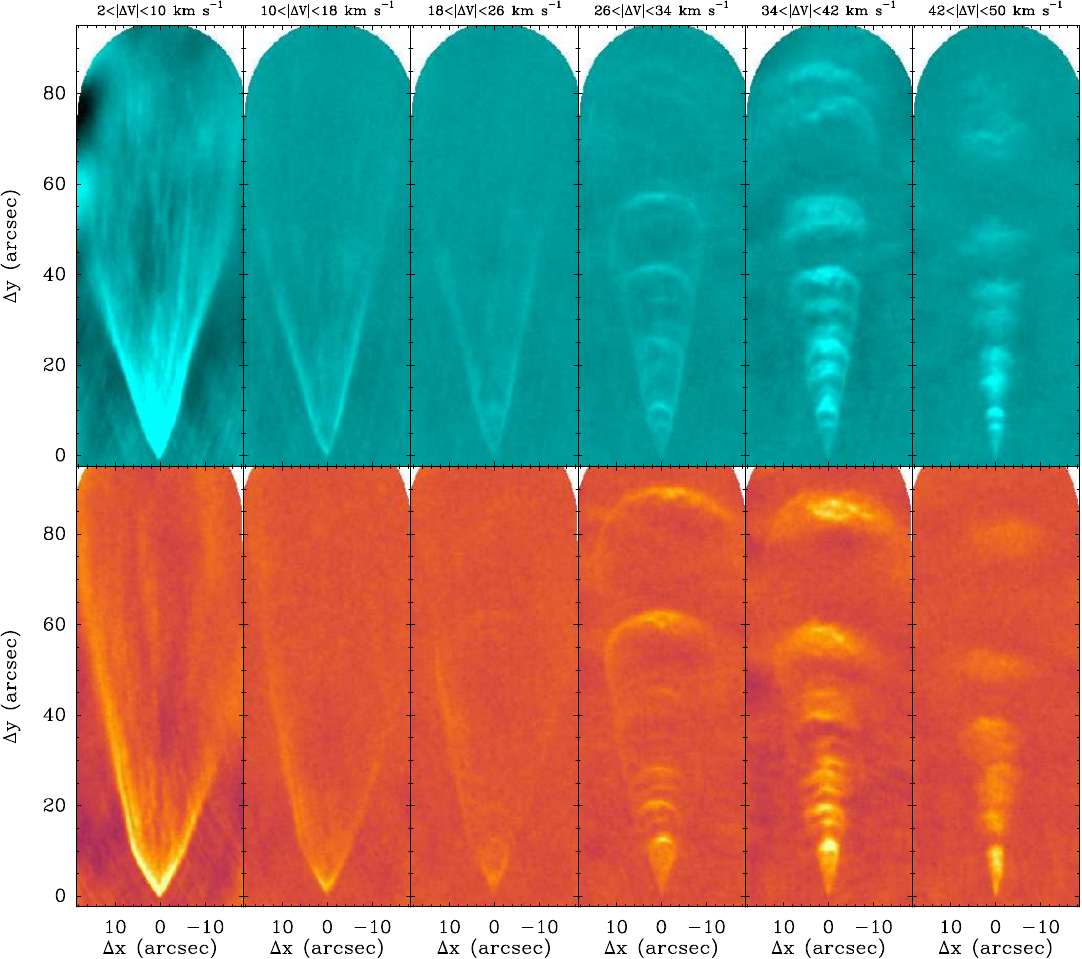}
        \caption{Maps of CO(2--1) intensity integrated over 8~km~s$^{-1}$
        intervals to illustrate the different velocity 
        regimes of the IRAS 04166 outflow. 
        For easier comparison, the maps of blueshifted
        emission (top) have been rotated clockwise by $30\fdg4$,
        while the maps of redshifted emission (bottom) have been rotated 
        by the same amount and reflected in the vertical direction.
        The interval of integration is given at the top and is measured
        with respect to the ambient LSR velocity of 6.7~km~s$^{-1}$.
        All maps are in linear scale with the extreme values 
        adjusted for maximum contrast. The position of IRAS 04166
        coincides with the origin of coordinates, and the blanked areas
        indicate regions not covered by the ALMA mosaic.}
        \label{fig_flow_rot}
\end{figure*}

The CO(2--1) maps of Fig.~\ref{fig_flow_4v} provide a first indication
of the outflow regular behavior and the strong symmetry 
between the two lobes. To study in more detail the velocity structure 
of the outflow, Fig.~\ref{fig_flow_rot} presents
a series of CO(2--1) velocity maps where the two lobes have been
oriented vertically so that the gas appears to flow upwards.
This has been done by rotating
the outflow clockwise by 30.4 degrees, which is the outflow position 
angle determined by \cite{san09}, and by reversing the y-axis in the red lobe.
To provide a finer view of the outflow kinematics,
the emission range of each lobe has been divided into six velocity intervals
of 8~km~s$^{-1}$.

The panels of Fig.~\ref{fig_flow_rot} provide further evidence 
of the regular behavior and symmetry of the two outflow components.
The three leftmost panels
cover the SHV regime and show that the slower outflow
gas is distributed in two conical lobes that have 
almost straight walls and the protostar at their
vertex. The maps also show that
the emission is significantly brighter toward the edges of
the lobes, suggesting that it exhibits
limb brightening caused by the 
gas being preferentially located in shells along the walls.

In addition, the opening angle of the conical lobes
decreases slightly with increasing velocity, 
which is a pattern previously seen in multiple outflows, both having a
jet-like EHV component, as in L1448 \citep{hir06}
and HH212 \citep{lee15}, and without having a EHV component,
as in L1551 \citep{mor87} and Mon R2 \citep{mey91}. 
This type of pattern suggests that the SHV emission traces
a shear flow composed of ambient gas that has been swept-up 
into shells by a faster outflow component that moves
along the center of the lobe. An alternative interpretation
in terms of a projection effect caused by gas moving
in a hollow cone with constant velocity can be ruled out. This model
predicts that after reaching its maximum opening angle, the
emission will start to collimate again 
with decreasing velocity due to the front-back 
symmetry of the gas moving in a cone. In reality, the 
emission of each lobe reaches its largest opening angle 
close to the ambient velocity and then disappears in the following
channel maps. This is consistent the physical decrease in velocity
expected for a shear flow, and not with a mere projection effect
expected for a hollow cone.

An additional feature within the low-velocity maps is the presence of
bright lines that emerge near the outflow base 
and extend almost vertically along the cavity. 
They are more prominent toward the blue lobe, but they can also be seen
in the red lobe with weaker intensity.
A possible mechanism for their origin is discussed in Section.~\ref{sect_consequences}.

\begin{figure*}
        \sidecaption
   \includegraphics[width=11cm]{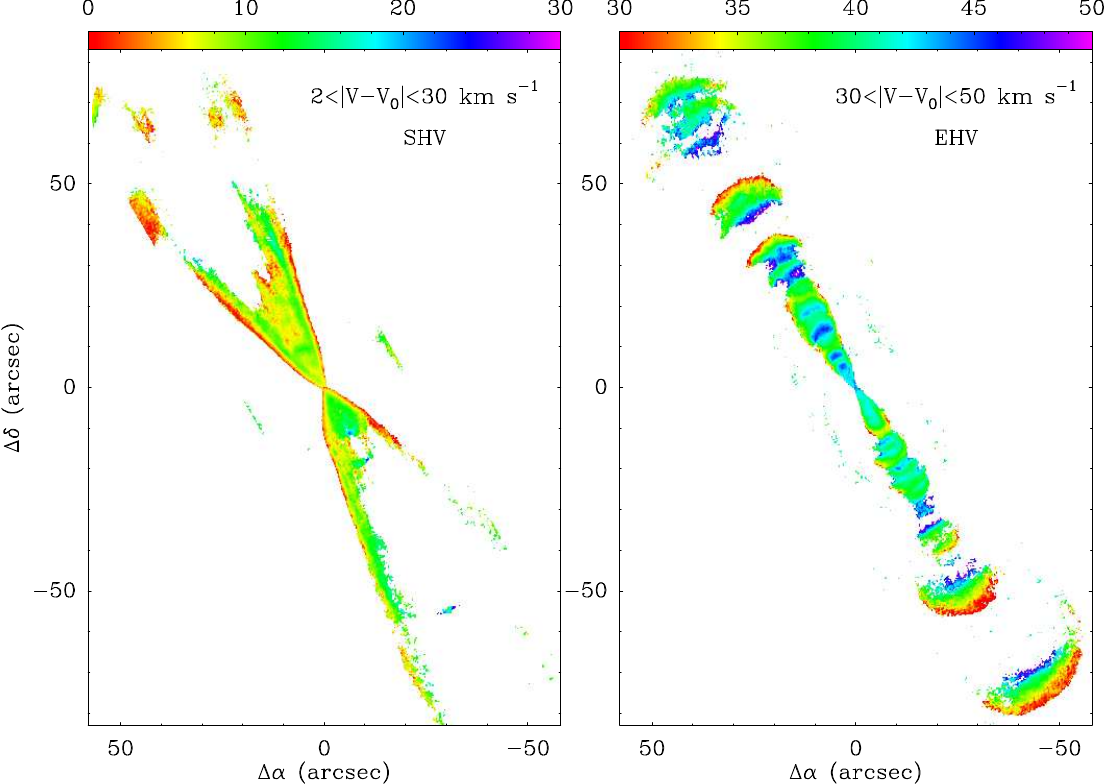}
        \caption{Maps of the absolute value of the 
        velocity centroid of CO(2--1) for the
        SHV (left) and EHV (right) regimes. IRAS 04166 is located
        at the origin of coordinates, and the color scales at the top are in units of km~s$^{-1}$. We note the shear pattern in the SHV regime and the 
        velocity oscillations along the outflow axis in the EHV regime.}
        \label{fig_flow_mom}
\end{figure*}

The three rightmost panels of Fig.~\ref{fig_flow_rot} approximately correspond to the
EHV regime, and show a very different distribution from that of the SHV gas.
At both blue- and redshifted velocities, the EHV emission delineates
a series of arcs concave with respect to 
the protostar that widen gradually with distance.
The position of these arcs matches well the position of the EHV peaks
previously seen by \cite{san09} and
\cite{wan14}, but the higher sensitivity of the ALMA images reveal now 
greater amount of detail.
In particular, the new ALMA data show that the EHV emission 
extends over the full width of the
SHV lobe, and that it merges smoothly with the side walls.
This behavior is best seen in the  $26<|\Delta V|<34$~km~s$^{-1}$
velocity range, which represents
the transition between the SHV and EHV regimes and therefore
contains emission from both velocity components. 
In these maps, the EHV arcs 
widen with distance from the protostar and  
follow the diverging walls of the SHV lobe,
suggesting that despite their different 
geometries, the SHV and EHV outflow components are physically connected. 
This is a first indication that the EHV gas
may play a role in the acceleration of the slower SHV outflow, an issue that is
further investigated in Sect.~\ref{sect_disc} with the help of PV 
diagrams.

Another notable feature of the EHV emission is its systematic
change of shape and position with velocity.
At the lowest EHV speeds ($26<|\Delta V|<34$~km~s$^{-1}$) the emission
is distributed in a series of arcs that are sharply defined and relatively thin. 
As the velocity increases, the emission of each arc becomes more diffuse
and shifts slightly toward the protostar, starting to fill the 
concave portion of the
arc. This behavior is more evident in the most distant 
EHV peaks of each lobe thanks to their larger separation and limited 
mutual overlap.
One EHV peak
in each lobe (the one near $\Delta y = 60''$) was studied by
\cite{taf17} using ALMA single-pointing observations, and its velocity pattern
was reproduced with a model where the gas is located along a
curved disk-like structure that is similar to a bow shock and 
where the gas is expanding laterally away from the outflow axis.
The new ALMA observations show now that this velocity pattern is common
to all the EHV peaks, strengthening the interpretation
that the EHV peaks represent internal bow shocks caused 
by a time-variable jet, as proposed by \cite{san09}.

To summarize of the main features of the outflow velocity field
we present in Fig.~\ref{fig_flow_mom} maps of the
CO(2--1) velocity centroid for the SHV and EHV regimes
using the actual orientation of the outflow. 
The left panel shows that
both the blue and red SHV lobes have a layered velocity field with the
slower gas (colored red) flowing along the outer boundary
of the lobe
and the faster gas (yellow to green) flowing interior to it.
In contrast, the EHV gas shown in the right panel presents multiple velocity 
oscillations along the outflow axis. These oscillations arise from the
already-mentioned
internal velocity structure of the EHV peaks, that have
the slowest gas located in the outer part of the arc (red color)
and the fastest gas in the inner part (purple). 
This pattern is more noticeable in
the furthermost EHV peaks partly due to their isolation but also partly due to the
choice of color scale. The gas in the inner peaks reaches slightly
higher velocities, as shown below with PV
diagrams, and the color scale has been chosen to emphasize the
velocity structure of the outer peaks.

\subsection{Connection between the SHV and EHV regimes}

As seen in the velocity maps of Fig.~\ref{fig_flow_rot}, 
the arced EHV emission extends over the full span of the SHV lobes and 
merges smoothly with their walls, suggesting that the two 
outflow components, despite their very different velocity signature, 
are physically connected.
To further investigate this possible connection between the EHV and SHV gas,
we turn our attention to the 
PV diagrams of the emission. 
These diagrams present the outflow velocity in one of their axes,
and are therefore better suited to study the continuity of the gas 
velocity field.

\begin{figure*}
        \sidecaption
   \includegraphics[width=12cm]{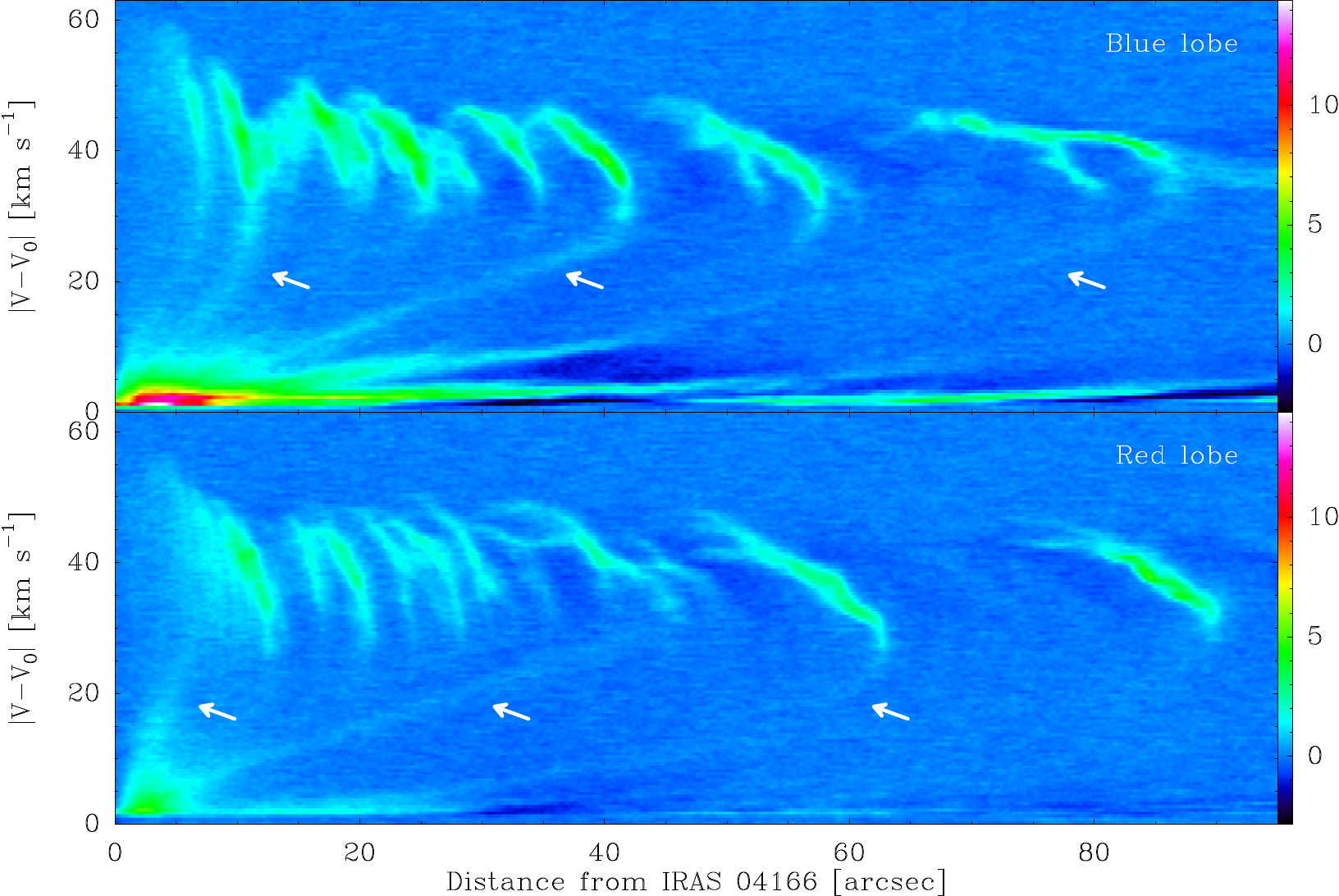}
        \caption{Position--velocity diagrams of the CO(2--1) emission 
        along the axis of the blue (top) and red (bottom) lobes of the 
        IRAS 04166 outflow. The black arrows indicate some of the 
        finger-like features that connect the EHV emission with 
        the origin of the diagram.
        The velocity scale is measured in absolute 
        value with respect to the cloud LSR velocity
        (6.7~km~s$^{-1}$), and the wedge scales on the right are 
        in units of K.}
        \label{fig_pv_along}
\end{figure*}

Figure~\ref{fig_pv_along} presents PV diagrams of the CO(2--1)
emission along the outflow axis
for both the blue (top) and red (bottom) lobes. These
diagrams have been generated as if a $4''$-wide slit had been placed
along the outflow axis, although the exact value of the slit width
has very little effect on the result. For easier comparison, 
the two diagrams represent the emission as a function
of the absolute velocity displacement with respect 
to the ambient cloud. Similar diagrams for the SiO(5--4) emission
are shown in Fig.~\ref{fig_pv_sio5}, although they are not further discussed due to 
their low sensitivity and their great similarity with those of CO(2--1).

As can be seen in Fig.~\ref{fig_pv_along}, the PV diagrams of the blue and red
CO(2--1) emission have very similar features. In each diagram, the emission 
from the EHV gas appears as a series of discrete components centered approximately
at $|V-V_0| \approx 40$~km~s$^{-1}$.
Each component presents a similar slanted orientation with 
a quasi-linear decrease in velocity from about 50~km~s$^{-1}$ to 30~km~s$^{-1}$
as the distance from the protostar increases.
This pattern corresponds to the systematic changes in the EHV emission seen in 
Fig.~\ref{fig_flow_rot} and the oscillations of Fig.~\ref{fig_flow_mom}.
Its repetition over the length of the outflow gives 
the PV diagram a characteristic sawtooth pattern 
that was previously identified by \cite{san09} and 
\cite{wan14}.

As discussed in \cite{san09}, the sawtooth velocity pattern of the EHV gas
matches the signature of a chain
of internal bow shocks driven by a jet with variable launch velocity.
In this type of jet, fast material launched at later times overtakes slower 
material that had been launched before, and the resulting shock 
compression ejects gas from the jet sideways.
As simulations show, the
sawtooth velocity pattern naturally arises in the PV diagram 
from the combined projection of 
the forward and lateral velocities of the ejected gas 
\citep{sto93,smi97,rab22,lor24}.

An expansion of laterally ejected gas from a jet also 
can explain the gradual flattening of the sawtooth pattern as 
a function of distance
to the protostar seen in Fig.~\ref{fig_pv_along}.
This behavior is expected from 
the gradual expansion of the internal bow shocks as they move
away from the protostar since the expansion spreads 
the lateral ejection over a larger spatial scale.
This effect is further explored in Sect.~\ref{sect_pv_model}
with the help of a ballistic model, but a similar behavior 
can be seen in the
simulations of \cite{wan19}, where the flattening is reproduced 
by the 
gradual expansion of spherical shells ejected by the protostar.

In addition to the sawtooth pattern,
the PV diagram of Fig.~\ref{fig_pv_along}
shows multiple linear features that connect
the EHV peaks with the origin of the diagram. 
The most prominent ones
are indicated with arrows in the figure, and 
can be seen almost symmetrically
in the blue and red lobes.
These features are referred to as fingers and have
not been previously seen in the IRAS 04166 
outflow probably because of their weak emission.
A close inspection of the PV diagrams suggests that there may be 
fingers connected to all EHV peaks, but that only a few  
are bright enough to be traced back
to the origin of the PV diagram. Higher-sensitivity observations
are needed to investigate the presence of additional fingers and
characterize their global distribution in the outflow.

The presence of fingers in the PV diagram suggests that each EHV peak
is followed by a trail of material at intermediate velocities 
that connects the EHV emission with the protostellar vicinity.
If the EHV peaks represent jet gas that has been ejected laterally
due to internal shocks,
the fingers would appear to 
correspond to the fraction of ejected gas that has been
left behind due to its interaction with the surrounding medium. 
Each EHV peak, therefore, likely represents the forward part of 
a shell of ejected gas, and 
since the fingers can be traced back to the protostellar vicinity,
the responsible lateral ejection has to have begun close
to the protostar.

This interpretation of the fingers 
is supported by the simulations 
of the interaction between a pulsating jet and its surrounding dense core
presented by \cite{rab22}.
As can be seen in their Fig.~9(c), the PV diagram of
material that has been accelerated by a pulsating jet 
presents a set of finger-like extensions that connect the origin
of the diagram with the peaks in the jet.
The presence of these fingers indicates that
linear momentum is being transferred from the EHV gas to the SHV regime,
slowing down the former and accelerating the latter.

If the fingers represent shells of gas expanding in the wake of an
EHV ejection, they are expected to appear as a series of ring-like structures
in PV diagrams perpendicular to the outflow, as shown by \cite{rab22}
in their Fig.~10.
Our PV diagrams of the IRAS 04166 outflow are significantly limited in sensitivity
due to the short integration time of the observations, but as shown in 
 Fig.~\ref{fig_pv_perp}, they do suggest the 
presence of ring-like cavities at intermediate
velocities. We therefore conclude that the main features in the
outflow PV diagrams, both parallel and perpendicular to the jet direction,
are best interpreted as resulting the from the lateral ejection of
jet material and the interaction of this material with the environment.
To further test this interpretation, we present in 
Sect.~\ref{sect_model} the results of a 
ballistic model of the interaction between
gas ejected laterally from a jet and the surrounding outflow
material. 
While highly simplified, our model allows exploring the velocity field
of the ejections needed to fit the PV diagram
along the IRAS 04166 outflow, and should be seen as a first step
toward a more comprehensible analysis 
to be carried out using numerical simulations.

\section{Discussion}
\label{sect_disc}

\subsection{Comparison to other outflows}
\label{sect_others}

The IRAS 04166 outflow is unusual for having an EHV component of
high symmetry
and collimation, but many of its other emission features 
have counterparts in less-extreme systems.
The presence of multiple outflow shells, for example, 
has been previously reported in different outflows,
especially when observed with ALMA.
\cite{lee15} found a very clean case of this 
type of geometry in the HH 212 outflow, where multiple nested shells 
can be seen in the CO emission
connecting the vicinity of the central source to the jet bow shocks
seen in H$_2$. 
Prominent nested shells
can also be seen in the CO emission of 
the highly collimated
CARMA-7 outflow mapped by \cite{plu15}, and similar
structures have been reported more recently by \cite{dev22}, \cite{omu24},
and \cite{liu25}. This frequent finding of discrete shell structures
superimposed on the large-scale emission from the bipolar outflow strongly
suggest that episodic ejection events play a prominent role in the 
mechanism of gas acceleration.

Another feature with counterparts in several outflows is the sawtooth
velocity pattern of the EHV gas.
\cite{hir10} found evidence for a similar pattern in the EHV gas
of the L1448 outflow,
and \cite{gav24} have recently presented a clean case of this type of
pattern in the jet-like outflow powered by BHR~71-IRS2.
Other outflows with  evidence for a sawtooth velocity 
pattern include those powered by 
MMS5/OMC-3 \citep{mat19}, HH270mm1-A \citep{omu24}, and
several sources mapped by the ALMASOP project \citep{dut24,liu25}.
In the case of HOPS 315, the pattern is accompanied by
finger-like extensions toward the origin of the
PV diagram like those seen in IRAS 04166, and
\cite{liu25} interpret them as resulting from the bubble 
structure of the unified outflow model of \citet{sha23}.
Additional evidence for finger-like extensions in an outflow PV diagram
has been presented by \cite{plu15} for the case of
CARMA-7, although
the low velocity of that outflow due to its high inclination
angle makes it difficult to discern whether the fingers are
connected to a sawtooth 
velocity pattern in the jet component.

Even outflows without evidence for an EHV component
have been found to present
finger-like features like those seen in IRAS 04166.
\cite{arc13} and \cite{zha19} have reported the presence of multiple 
finger-like shells in 
the PV diagrams along the HH 46/47 outflow.
Additional evidence for PV fingers has been presented by \cite{non20}
in several outflows of the W43-MM1 protocluster, 
\cite{vaz21} in the outflows of the low-mass Lupus cloud,
and \cite{dev22} in the outflow of DG Tauri B.
In all these cases, the fingers have been interpreted as 
resulting from some type of episodic event in the outflow driving
agent, although the nature of this agent has been variously 
assumed to be
a wide-angle component \citep{arc13,zha19,vaz21},
a collimated jet \citep{non20},
or an MHD disk wind \citep{dev22}.

The finding of shells, sawtooth patterns, and PV fingers in outflows 
that are at different evolutionary stages and located
in different environments
suggests that these features may represent a frequent 
component
of the outflow phenomenon that needs to be better understood.
In this respect, the IRAS 04166 outflow, with
its simple geometry and high degree of symmetry, offers an ideal
target to model the origin of these different features
and to test
whether they could result from the interaction of
a time-variable jet with its surrounding medium.
Given the limited number of simulations of realistic jets available, 
we rely for our analysis on a
highly idealized model of the interaction between
gas ejected from a pulsating jet and its surrounding medium
that is presented in the next section.

\subsection{A toy model of the lateral ejection from a jet}
\label{sect_model}

\begin{figure*}
\centering
   \begin{minipage}[b]{0.4\textwidth}
      \includegraphics[width=0.95\textwidth]{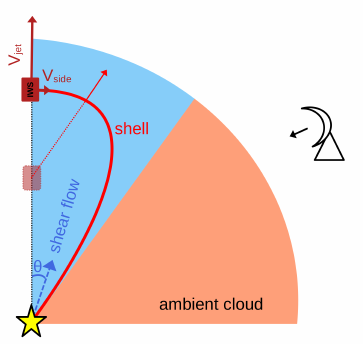}
   \end{minipage}
   \begin{minipage}[b]{0.4\textwidth}
        \includegraphics[width=0.95\textwidth]{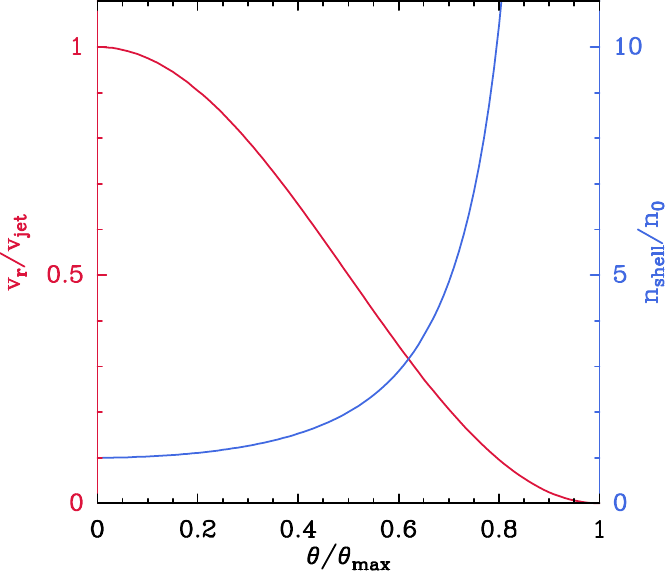}
   \end{minipage}
\caption{{\it Left:} Geometry assumed for the simple model of
        lateral ejection from an internal working surface (IWS) 
        presented in Sect.~\ref{sect_model}
         (only one outflow quadrant is shown).
         The IWS moves along the y-axis with a constant $v_{\mathrm{jet}}$
         velocity and ejects gas with a $v_{\mathrm{side}}$ 
         perpendicular component. The ejected gas interacts with
         a surrounding shear flow and forms a shell that surrounds
         the jet axis (red solid curve). 
         The plot shows the current location of the IWS (solid) and 
         its position approximately half way along its path (dotted).
         The red dotted line represents the approximate 
         trajectory between time of ejection and current time.
         The model predicts the expected PV diagram from an observation 
         made at an arbitrary angle from the jet axis.
         {\it Right:} Velocity and density profiles
         as a function of angle from the jet axis used to simulate
         the outflow shear flow. They correspond to Eqs. 1 and 2 in the text.}
        \label{fig_model}
\end{figure*}

As we have seen, the
simulations of pulsating jets by \cite{sto93}
and \cite{rab22} predict features in the emission maps and PV diagrams
that resemble those found in the IRAS 04166 outflow.
Unfortunately, these simulations cannot be easily compared 
to our ALMA data since they assume specific 
jet and cloud conditions that do not necessarily coincide with those of 
the IRAS 04166 outflow.
To further investigate the origin of the 
features seen in the IRAS 04166 outflow, 
we present a highly simplified model of 
the interaction between gas ejected sideways by an internal
shock in a jet and the surrounding outflow material.
Given its simplicity, we consider this model only as a first step
toward a more realistic description of the interaction between a
jet shock and its surrounding flow.

Fig.~\ref{fig_model} presents a schematic view of how the model
simulates the evolution of a single lateral ejection produced by 
an internal shock in a jet.
The model assumes that the jet propagates along the y-axis,
and that due to some type of velocity variability, it
develops an internal working surface (IWS)
of high compression (\citealt{rag90}).
This IWS moves along the y-axis and ejects
gas perpendicular to its motion.
The ejected gas interacts with the surrounding outflow material, and 
due to the ensuing compression and possible heating, the combined
ejected plus swept up material produces a visible signal 
in the outflow emission. 
The goal of the model is to follow the path of the ejected gas as 
it moves through the surrounding medium, and to calculate
how much mass is swept up in the process assuming conservation of 
linear momentum.
To keep the calculation as simple as possible, the ejected gas
is treated 
as a set of independent particles that move away from the 
jet beam with a given perpendicular velocity in addition to
their initial jet velocity.

A critical ingredient of the model is the velocity field assumed for
the gas that surrounds the jet. This gas has been accelerated 
by multiple previous ejections, and therefore has a potentially 
complex kinematics that can only be treated in a highly idealized way.
As mentioned in Sect.~\ref{sec_vel_str}, the velocity of the
SHV component in the IRAS 04166 outflow
decreases gradually toward the walls of the lobes and
follows a shear-flow pattern.
Similar shear-flow patterns are predicted by
the simulations of jet-driven outflows by \cite{rab22} as a result
of the interaction between multiple shocks from successive 
jet ejections (their Sect.~5.2), and by the model of a 
wide-angle wind of \cite{sha06,sha20,sha23} 
as an intrinsic property of the outflow
(see also \citealt{lia20}).
Following these theoretical results, we model
the gas around the jet as having a shear-flow pattern in which
the material flows radially from the protostar with a speed that depends on
the angle $\theta$ with respect to the jet direction (Fig.~\ref{fig_model}).
For continuity, the flow velocity is assumed to match 
the jet velocity along the y-axis ($\theta = 0$)
and to decrease gradually until reaching a zero value
at $\theta_{\mathrm{max}}$, which represents the boundary
between the outflow and the ambient cloud.
Since the model is highly idealized and its goal is to
reproduce the general behavior of the EHV emission, we 
choose the following
simple velocity field that satisfies the above constraints:
\begin{equation}
v_{\mathrm{r}}(\theta)= v_{\mathrm{jet}} \;  \cos^2 \left(\dfrac{\pi}{2} \dfrac{\theta}
{\theta_\mathrm{max}}\right). 
\end{equation}
Here $v_{\mathrm{r}}$ is the radial velocity of the gas, 
$v_{\mathrm{jet}}$ is the jet forward velocity, and $\theta_\mathrm{max}$
is the outflow half opening angle (Fig.~\ref{fig_model}).

To complete the description of the gas surrounding the jet
we need to model its density profile. 
The maps of Fig.~\ref{fig_flow_4v} show the SHV component
as a limb-brightened conical shell,
indicating that its density must
increase toward the outer walls of the outflow.
A similar gas distribution is predicted by the simulations of
\cite{rab22} (their Sect.~5.2), so favoring again a simple expression, 
we choose a density profile of the form
\begin{equation}
n_{\mathrm{shell}}(r,\theta) = n_0(r) \;  \cos^{-2} \left(\dfrac{\pi}{2} \dfrac{\theta}
{\theta_\mathrm{max}}\right),
\end{equation}
where $n_0(r)$ is a fiducial density that in principle could depend on 
the distance to the protostar $r$. This dependence is likely flatter than
the $r^{-2}$ expected for conservation of mass since the outflow 
consists of ambient accelerated gas that is gradually being entrained along the
cavity walls (plus an additional contribution from ejected jet material).
To test the sensitivity of our results to the assumed radial dependence of
the density profile, we run models of the best-fit solution (described in the next 
section) using $n_0(r)$ factors with $r^{-2}$, $r^{-1}$, and no radial dependence.
The resulting PV diagrams (shown in Fig.~\ref{fig_dens_test}) indicate that 
the radial dependence of the density profile plays a minimal role in the 
solution so, for simplicity, we take $n_0(r)$ as a constant
value.

\begin{figure*}
   \includegraphics[width=0.9\hsize] {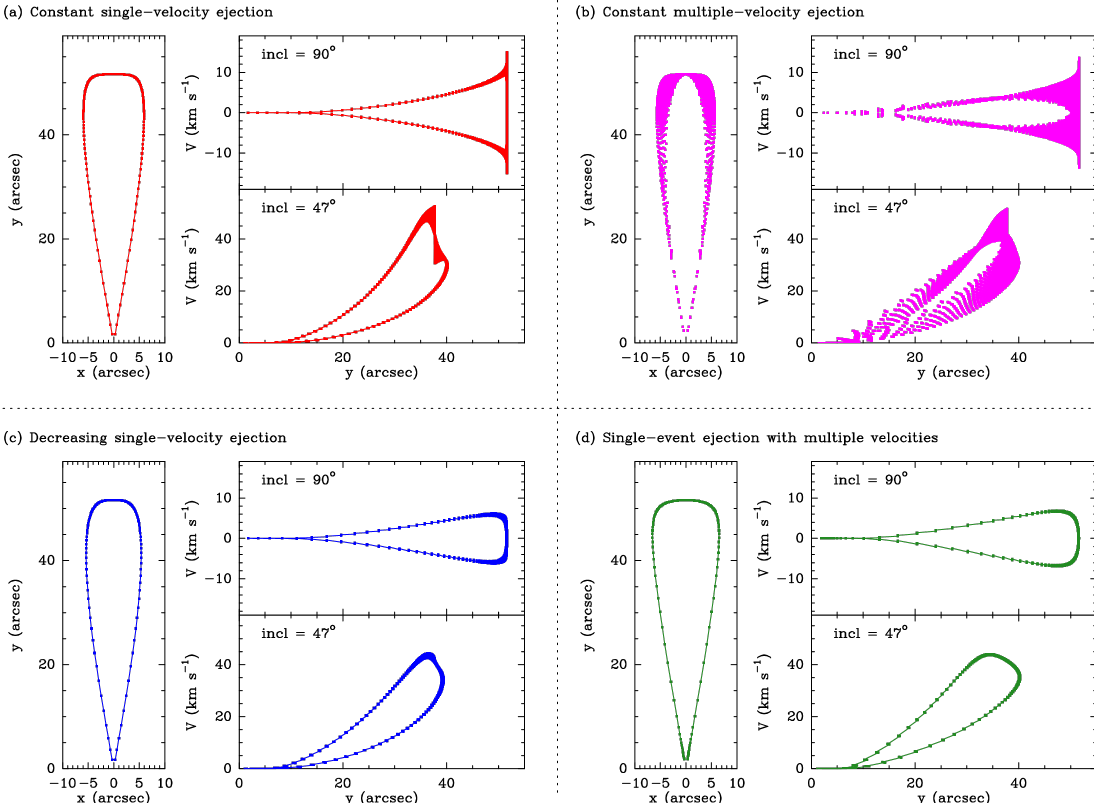}
        \caption{Summary of results from modeling 
        four different modes of lateral ejection
        from a jet.
        Each mode is represented by a quadrant, and is labeled 
        with its kinematic characteristics and color-coded 
        for easier reference. In each quadrant, the vertical diagram
        represents a map of the ejected gas, and two horizontal
        diagrams are position--velocity plots for inclinations of
        $90^\circ$  for a jet moving in the plane of sky and 
        $47^\circ$  for expected inclination of the IRAS 04166 outflow (see text for details).}
        \label{fig_model_comparison}
\end{figure*}

Having defined the environment where the ejection propagates, the 
model follows a set of particles of given initial mass
and density that are ejected from the IWS as it moves at constant velocity 
along the jet axis.
The trajectory of each particle is determined at consecutive time steps
by requiring conservation of the combined linear momentum of the ejected gas and the
swept-up material.
At a chosen time, the positions and velocities of the particles 
can be used to generate the PV diagram that results from an
observation from an arbitrary angle with respect to the jet axis.
Given the assumptions of the model, we refer to it as 
the LEAF (Lateral Ejection onto an Angle-dependent Flow) model.
We note that given its simplicity, the model cannot be used to investigate
the nature of the jet-like wind (pure jet, disk wind, or X-wind) 
but only its effect on the surrounding medium.

\subsection{Different modes of lateral ejection and their signature in the PV diagram}
\label{sect_model_results}

As a first application of the LEAF model, we 
investigated how the properties of a lateral ejection
affect the PV diagram.
For this, we run different cases varying the two main
parameters that seem to
control the evolution of the ejected gas in the model:
the rate at which the IWS ejects material laterally 
and the distribution of velocities in the ejected gas.
To explore the effect of the ejection rate, we compared 
cases where the IWS ejects material continuously as it propagates 
along the jet axis with cases where the material is ejected
in a single event. To investigate the effect of the distribution
of ejection velocities, we compared cases where all the ejected gas has
the same lateral velocity with cases where the ejected gas contains a
mix of lateral velocities. 
To simplify the discussion, here we present
the results from four types of ejection
that reproduce some of the features seen in the PV diagrams of 
the IRAS 04166 outflow: 
(a) continuous ejection with a single lateral velocity,
(b) continuous ejection with multiple lateral velocities, 
(c) continuous ejection with a single decreasing velocity,
and (d) single-event ejection with multiple lateral velocities.

Figure~\ref{fig_model_comparison} summarizes the main results of
our model exploration by presenting the output of each case in one quadrant.
The elongated diagram on the left of each quadrant represents the
unprojected map of the ejected material, and 
the two plots to its right represent PV diagrams along
the outflow axis. The top one corresponds to an inclination of
$90^\circ$ (jet moving in the plane of the sky)
and the bottom one to an inclination of 
$47^\circ$, which is the value estimated for IRAS 04166
by the eDisk project from a fit to the disk emission
\citep{oha23,phu25}. We prefer
this value to the similar one of $52^\circ$ 
derived by \cite{taf17}
because the disk measurement is likely more accurate than
an estimate based on the outflow emission, although the exact
choice of angle has little influence in our analysis.

Since the LEAF model solves the ejection of gas in a single plane, 
we  simulated the three-dimensional structure of the outflow
by combining multiple LEAF solutions, each representing the
ejection of gas in a different plane around the jet axis.
Using this set of
LEAF solutions, we generated the PV diagram
expected from an observation along the jet axis
with a slit 4 arcsec wide
like that used with the IRAS 04166 
outflow in Fig.~\ref{fig_pv_along}.
In all models, we assumed a jet velocity of 61~km~s$^{-1}$ 
for consistency with \cite{taf17}, and we followed the ejection 
of jet material for
650~yr, which is the time needed to match the position
of one of the IRAS 04166 EHV peaks in the PV diagram
(Sect.~\ref{sect_pv_model}).

As can be seen in Fig.~\ref{fig_model_comparison}, the 
panels that represent the maps of 
the ejected material for all four ejection cases show
similar distributions that broadly resemble the individual 
outflow shells of IRAS 04166 seen in Fig.~\ref{fig_flow_rot}. 
While this good match reinforces the lateral-ejection interpretation 
of the shell geometry, it also
shows that the spatial distribution of the ejected gas 
can be fit using different ejection options, so it does not
provide a strong constraint on the gas kinematics.

In contrast with the maps, 
the PV diagrams of Fig.~\ref{fig_model_comparison}
show a variety of shapes that suggests they
are more sensitive to differences in the mode of ejection.
Our first model (top left quadrant) assumed that the IWS 
moves along the jet axis and simultaneously ejects material laterally
at a constant rate and with a single velocity of 15~km~s$^{-1}$. 
This velocity choice is in line with the estimate from
\cite{ost01}, who predicted that in the bow shock of a protostellar jet 
the gas is heated to $10^4$~K and ejected with a speed of 
$\sim\!\!\sqrt5$ times the sound speed.
If the jet is moving in the plane of
the sky ($90^\circ$ inclination), the predicted PV diagram presents
a trumpet shape with two branches that diverge with distance
from the central source and are connected by a straight line
at the IWS position. This connection arises from jet gas that
moves away from the jet axis at different angles with respect to the
line of sight and therefore is projected inside the slit 
at intermediate velocities.
The trumpet shape resembles that predicted by 
more realistic models of a single bow shock, like those
of \cite{ost01} (their Fig. 3) and \cite{lee01} (their Fig. 5).
These models also assumed a constant lateral ejection velocity, so their
agreement with our predicted PV diagram suggests that despite its 
simplicity, the LEAF model
captures the basic kinematic properties of the of lateral 
ejection.
For an inclination angle of $47^\circ$, the LEAF model predicts
a connection between the two branches of the PV diagram 
that has a complex bow-shape
structure and differs significantly from the smooth linear 
distribution
seen in the IRAS 04166 PV diagram (Fig.~\ref{fig_pv_along}).
We therefore conclude that a lateral 
ejection model having a constant single velocity 
fails to reproduce the observed PV diagram.

As an alternative model,
we explored a case where the IWS moves along the jet axis and 
simultaneously ejects material with
a mixture of lateral velocities. 
For simplicity, we assumed that the mixture is
uniformly distributed between zero and 15~km~s$^{-1}$
(the velocity of the single-velocity model),
and present the results in the top right quadrant of 
Fig.~\ref{fig_model_comparison}.
As can be seen, the PV diagram for an inclination of $90^\circ$ 
consists of a cloud of points whose outer
envelope coincides with the single-velocity model,
but is now filled with points toward the end of the diagram.
For an inclination of $47^\circ$, the PV diagram
resembles that of the single-velocity model,
but has a broader distribution of points at high velocities
that contrasts with the thin EHV features seen in the IRAS 04166
PV diagram. 
A multiple-velocity model, therefore, also fails to
reproduce the observations.

The failure of models with a constant rate of ejection reinforces
the analysis of \cite{taf17}, who found that to
model the emission from two EHV condensations, the lateral velocity
of the gas had to increase linearly with distance from 
the jet axis.
To achieve this, two possible mechanisms were suggested:
(i) the velocity of the lateral
ejection decreases linearly with time, or
(ii) the ejection occurs during a single event that generates
multiple lateral velocities, and 
the material orders itself into a Hubble flow as it expands sideways.
Our next two models explored these two options.

To model an ejection with a decreasing lateral velocity
we assumed that the IWS starts ejecting gas laterally at 15~km~s$^{-1}$
when close to the protostar, and that the lateral velocity
decreases linearly as the IWS advances, until it reaches zero
at the time the system is observed.
As can be seen in the bottom left quadrant of 
Fig.~\ref{fig_model_comparison}, 
the resulting PV diagrams present a smooth distribution
that for an inclination of $47^\circ$ matches reasonably well the 
shape of the individual 
features of the IRAS 04166 PV diagram. In this respect, a model
with decreasing 
ejection velocity seems to provide a reasonable fit to the ALMA
observations.
A caveat, however, is that the model requires that 
by the time when the system is observed, the ejection velocity has 
decreased from its initial high value to zero, or
otherwise the PV diagram would include a 
bow-tie structure similar to that seen in our first model.
Since none of the EHV peaks in 
the PV diagram of the IRAS 04166 outflow presents this type of
structure, the condition of having reached zero velocity 
has to be satisfied by all the outflow ejections, no matter
how close they are located from the protostar (and therefore how 
young they are).
This suggests that the ejection velocity has to decrease very
rapidly from its initial value to zero. 

A lateral ejection whose velocity decreases rapidly
to zero is practically equivalent to an ejection that 
releases the gas instantaneously with a
distribution of velocities between a maximum value and zero. This
type of ejection
was the second alternative proposed in \cite{taf17} 
to fit their observations, and
is the final ejection case that we explored
using the LEAF model.
To follow its evolution, we assumed that
the ejection takes place when the IWS is at a very small
distance from the protostar, which we took as
the size of the first time step 
to avoid possible divergences 
caused by the zero radius at the origin.
This instantaneous
ejection was assumed to release gas that has a uniform mix of
lateral velocities between zero and 15~km~s$^{-1}$.
As shown in the
bottom right quadrant of Fig.~\ref{fig_model_comparison},
the predicted PV diagram for an inclination of
$47^\circ$ reproduces well the
main characteristics of each EHV peak in the IRAS 04166 
outflow: a fast regime where the
velocity decreases linearly 
with distance from the protostar and finger-like extensions that 
connect the fastest emission to the origin of the PV diagram.
A single-event ejection containing
multiple lateral velocities therefore provides the best fit to the 
the PV diagram of the individual EHV peaks in the IRAS 04166
outflow.

While the results of our simple LEAF model need confirmation
using more realistic simulations, they already provide some clues
on how the different features of the PV diagram may 
arise from the lateral ejection of jet material.
For the case of a single-event ejection, the LEAF model suggests
that the ejected gas leaves the IWS moving with
a combination of a forward jet velocity and a lateral
component that lies in the range from zero to about 15~km~s$^{-1}$. 
The gas ejected with the largest lateral 
velocities moves rapidly away from the jet axis and sweeps
gas of increasing density from the surrounding shear flow, slowing
down in the process. This fraction of gas gives rise to the
PV fingers, which are therefore composed of a mix of original 
ejected material and swept-up surrounding gas.
In contrast with this rapidly decelerated gas, the fraction of material 
ejected with a small lateral velocity component
moves along the jet direction
and suffers less deceleration
from its interaction with the surrounding shear flow due
to the combined effect of the high velocity of the shear flow close
to the jet axis and its lower volume density. As this gas moves
forward, it also spreads laterally and acquires a Hubble-law 
distribution with respect to the jet axis, giving the head of the
PV diagram its characteristic slanted shape 
This fraction of gas corresponds to the EHV
component, and is distributed spatially as a flattened,
curved bow shock. 
In this respect, the description of the EHV component inferred from
the LEAF model is consistent
with the more-ad hoc parameterization
used in \cite{taf17} to reproduce the velocity maps
of the EHV emission using a parabolic distribution of expanding gas.
The advantage of the new LEAF model is that
in addition to fitting the kinematics of the EHV gas
it automatically reproduces the finger features that connect the 
EHV emission with the origin of the PV diagram.

\subsection{Simulating the multiple ejection events of the IRAS 04166 PV diagram}
\label{sect_pv_model}

\begin{figure}
   \includegraphics[width=\hsize] {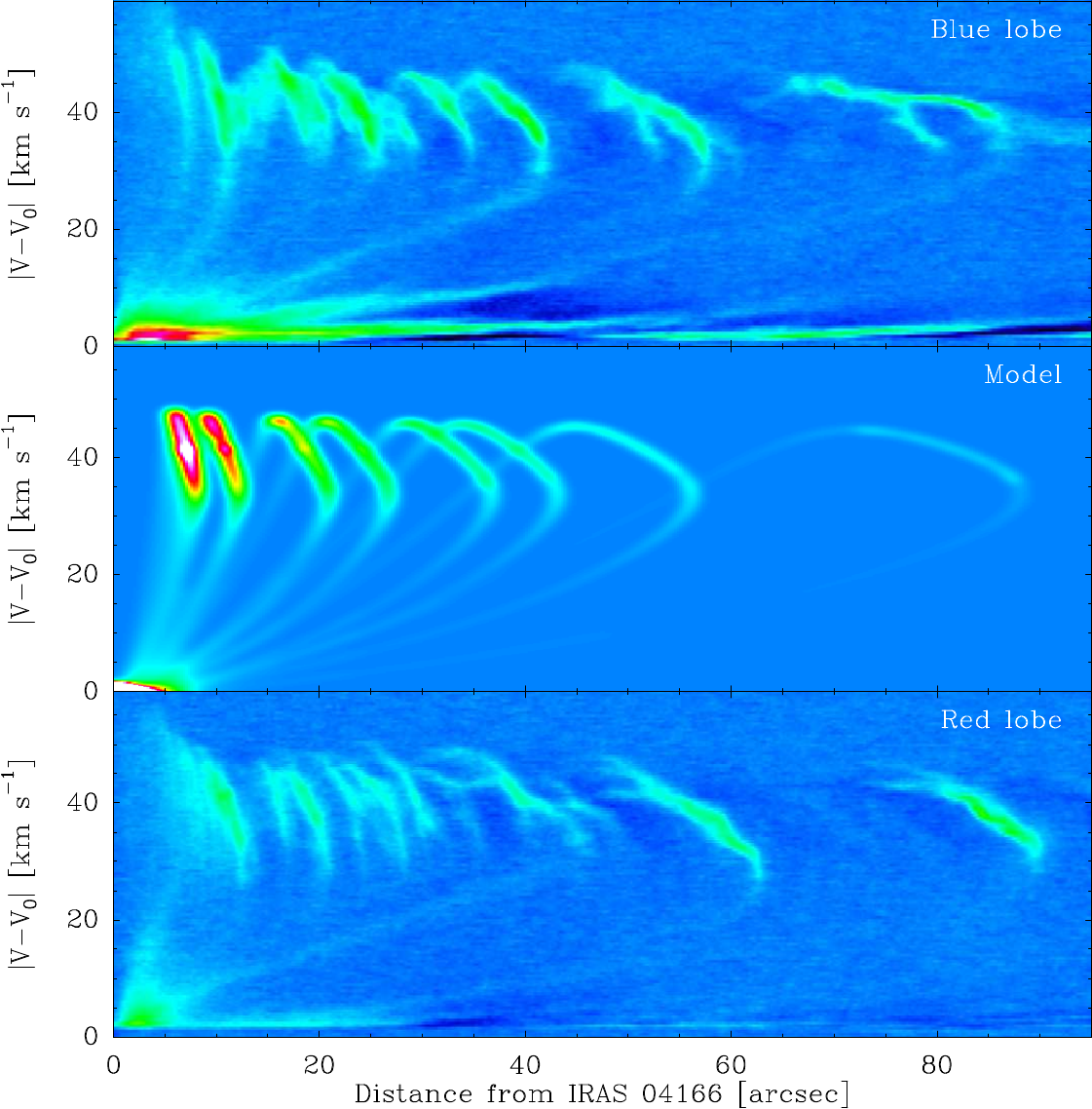}
        \caption{Comparison between the PV diagram from a 
        multi-ejection model (middle) and the ALMA observations
        of the IRAS 04166 outflow
        (as shown in Fig.~\ref{fig_pv_along}).
        The intensity scale of the model is linear but arbitrary 
        since it is simply scaled from the amount of ejected plus
        swept-up gas.}
        \label{fig_pv_vs_model}
\end{figure}

Having reproduced the shape of the individual PV-diagram features
assuming an instantaneous ejection, we attempted to
model the full EHV emission of the IRAS 04166 
outflow by simulating a sequence of successive ejections that travel 
along the jet axis. For this we 
run multiple instances of the LEAF model using the same
physical conditions for the jet and the shear flow, and only varied the
time at which each ejection was emitted.

From the PV diagram of the blue lobe, which shows a brighter and cleaner pattern,
we estimated that there are at least eight
separate ejections propagating in the
region mapped with ALMA.
This number is higher by one than estimated
by \cite{san09} from their lower-sensitivity data, although the 
exact value  
is still uncertain due to the complexity of the emission near the
protostar, where some of the ejections lie very close to each other and
may be interacting among themselves. This uncertainty is not critical
for our model since we did not aim to fit exactly the emission but to show
how its main features can be understood as a resulting from the 
propagation of multiple ejection events.

To determine the age of each EHV peak, we followed with the LEAF model
the evolution of the ejected gas from its launching point 
near the protostar to its current position.
As before, we assumed a constant jet velocity
of 61 km~$^{-1}$
and an inclination with the line of sight of $47^\circ$.
Since the jet velocity and the shear-flow conditions are 
assumed to be the same for all the ejections, 
the sequence of the eight model ejections can also be seen as a time
series in the propagation of a single generic ejection.
By fitting the position of their EHV heads, we found that 
the eight episodes
need to have the following approximate ages:
120, 180, 310, 400, 500, 650, 850, and 1350 yr.

While the above values do not indicate a strictly periodic pattern, 
we note that the first six episodes have relative separations close to
100 yr, after which the time between episodes increases significantly. 
Similar patterns of increasing separation between ejection events 
are seen in some HH jets \citep{rei01},
and may arise from 
the gradual fading of some ejections as they move away from the protostar
or from the dissociation of CO in particularly strong shocks.
If this is the case in IRAS 04166, 
our results suggest that the outflow lateral 
ejections have a typical time scale of 
around 100~yr. This time scale most likely 
reflects some type of variability in the accretion process
since the central source appears to be single \citep{phu25}.

Although the LEAF model simulates the kinematics of the ejected gas
and predicts its location on the PV diagram, it does not 
solve the equation
of radiative transfer, so it cannot be used to derive
line intensities. We can still estimate
an approximate level of EHV
emission by using the distribution of moving mass
and assuming that the CO(2--1) emission is optically thin. 
This assumption is consistent with 
our lack of detection of $^{13}$CO(2--1) in the 
EHV regime and with the quantitative analysis of the emission
by \cite{taf10}.
Using this approximation, and convolving the resulting 
PV diagram with a small kernel to simulate a finite 
duration of the ejection event and the jet velocity dispersion,
we obtained the PV diagram shown in 
the middle panel of Fig.~\ref{fig_pv_vs_model}.

As can be seen in Fig.~\ref{fig_pv_vs_model}, a model with
multiple instantaneous
ejections reproduces most of the features seen in the PV diagram,
including the sawtooth appearance of the EHV emission and the 
presence of low-velocity fingers connecting the EHV emission with the
origin of the diagram. It also reproduces the
gradual flattening of the sawtooth pattern as the ejections move away from
the protostar. This behavior results from the lateral expansion of the 
ejected gas, which
stretches its emission along the spatial axis of the PV diagram
while it maintains a constant velocity spread 
(see \citealt{wan19} for a similar interpretation in terms of 
expanding spherical shells).
Another feature seen both in the data and the model is the relatively 
weak emission  
of the low-velocity fingers. 
Unfortunately the ALMA observations are not sensitive enough to reveal 
the full
complexity of the overlapping emission from the fingers that is expected 
in the model, so it is not possible to 
use these features to further constrain the interaction between the
EHV and SHV gas.

While not directly predicted by our model, the interpretation of the 
EHV gas as a lateral jet ejection that propagates into the surrounding
medium can also help explain the bright trails seen to radiate 
from the inner outflow and extend almost vertically 
in the low-velocity maps of Fig.~\ref{fig_flow_rot}. 
These trails could represent 
lateral ejections that due to an anisotropy in their launching
are impacting the surrounding gas in a preferential direction
as they propagate along the outflow. Further observations of 
these features are needed to better constrain their 
velocity pattern.

\subsection{Evolution and detectability of the unperturbed jet}
\label{sect_consequences}

The interpretation of the EHV emission as representing
jet material that has been ejected laterally
by internal shocks
raises the question of whether there is any evidence in the
data for jet gas that has not yet undergone a
shock interaction. 
Graphic depictions and models of jet-driven outflows
often show a jet component traveling
between consecutive internal working surfaces
that has not yet encountered an internal shock
(e.g., Fig.~1 in \citealt{rag93a} or Fig.~19 in \citealt{dev22}). 
The IRAS 04166 data, however, do not show any evidence 
for such an unperturbed jet component, and the only emission features
that could potentially be associated with a pristine jet are
the two unresolved EHV peaks seen toward the protostar
vicinity in SiO(5--4) (see Fig.~\ref{fig_ehv_center}).
All EHV emission at further distances from the protostar
displays the characteristic sawtooth velocity pattern that we
have associated with laterally expanding jet gas.

While it is still possible that an unperturbed jet component 
remains undetected because it has an atomic composition,
this seems unlikely since
it would require that the jet gas becomes molecular
almost instantaneously after being ejected by the shock.
It seems more likely that the lack of detection of unperturbed jet material
between the EHV peaks arises because most (or all) the original
protostellar jet has been converted into laterally expanding gas. 
This interpretation agrees with the
LEAF model analysis, which requires 
that the ejection of gas from the jet 
occurs almost instantaneously in the vicinity of the protostar.
It also helps to understand the lack of significant emission  
of optical and IR shock tracers along the jet axis, which was
mentioned in the introduction.
Any shocks responsible for the lateral ejection of jet gas
must have occurred in the protostellar vicinity, which is highly
extincted.

Whether the above interpretation agrees
with published simulations of time-varying jets is less clear, 
although we note that \cite{sto93} state that the jet beam in their
simulations is rapidly depleted (``starved'') 
due to the lateral ejection of material by the internal shocks.
A detailed study of the survival of unperturbed jet material
using numerical simulations would be highly desirable to
clarify our understanding of jet propagation and evolution.

To summarize, the picture that emerges from 
the IRAS 04166 data 
is one where a protostellar jet is ultimately
responsible for the production of the EHV component
via internal shocks.
These shocks, in turn, lead to a rapid destruction
of the jet and its conversion into a series of forward-moving,
but also laterally expanding shells of gas.
Whether a similar scenario is applicable to other outflows
still needs further investigation, but the presence of finger-like
structures in other systems
suggests that this is a realistic possibility. Since many of these outflows
lack an EHV component in molecular lines, we have to assume
that the gas ejected from the jet in those cases 
has an atomic composition,
and that the finger signatures in these outflows
represent ambient material that has been accelerated by 
the atomic component. A more systematic investigation of the 
frequency and properties 
of finger structures in other outflows is necessary to test this
scenario.

\section{Conclusions}

We  observed the IRAS 04166 outflow with the 12m
and Compact Arrays of ALMA using the Band 6 receivers
and focusing on the CO(2--1) emission. We  also developed
a simplified model of the evolution of gas ejected laterally
from a jet by an internal working surface, and we  used this model
to interpret the PV diagrams of the CO(2--1) emission. 
From this work, we have reached the following main conclusions:

1. In agreement with previous observations, we find that the IRAS 04166
outflow comprises two regimes that we refer to as 
the SHV and EHV components. The SHV component appears in the maps
as a pair of limb-brightened conical shells where the gas
follows a shear-flow velocity pattern.
The EHV component lies internal to the SHV shells and appears
in the maps as a series of arcs similar to bow shocks. Each
of these arcs presents a similar velocity pattern 
with the faster gas located closer to the protostar and the slower 
gas at larger distances.

2. The new ALMA data reveal a previously unseen connection between the
SHV and EHV components of the outflow. The velocity maps show that the
EHV arcs span the full width of the SHV shells and merge smoothly
with their walls. PV diagrams of the emission along the outflow axis show
finger-like features that connect the EHV emission with the origin of
the diagram. These features suggests that as it moves forward,
the EHV-emitting gas leaves behind a trail of gas that has decelerated
due to its interaction with the surrounding SHV component.

3. A simple model of the lateral ejection of gas from an internal working
surface can reproduce both the velocity pattern of the EHV gas and the
presence of the finger extensions. For this to happen, the gas 
needs to have been ejected almost
instantaneously close to the protostar. According to this model, the 
EHV gas and the finger extensions represent different parts of the
ejected gas that started their expansion with different amounts of
lateral velocity. 
Using this model, it is possible to reproduce the PV diagram of
the outflow assuming that there have been eight ejections 
with a typical separation of a hundred years.

4. If the EHV emission represents material that
has been ejected from a jet near the protostar, no emission
in the data seems associated with a component of the jet
that has not yet undergone shock interaction
(with the possible exception of unresolved emission toward the
protostar). The data therefore suggests that the jet is fully disrupted
by the internal shocks
early on in its propagation, and that any later interaction
between the jet material and the surrounding gas occurs through
the expansion of the laterally ejected gas. In the IRAS 04166 outflow
this interaction is manifested through the finger structures
in the PV diagram.
Similar fingers in other outflows may also result from
lateral ejections from a jet, although in many cases the
ejected gas must be atomic given the lack of molecular EHV
emission. Further analysis of finger structures in other outflows
is needed to test this scenario.

\begin{acknowledgements}
We thank John Bally for his insightful referee's reports that helped us 
refine this work.
M.T. acknowledges partial support from project PID2019-108765GB-I00
funded by MCIN/AEI/10.13039/501100011033.
D.J.\ is supported by NRC Canada and by an NSERC Discovery Grant.
This paper makes use of the following ALMA data: ADS/JAO.ALMA\#2021.0.00575.S. 
ALMA is a partnership of ESO (representing its member states), NSF (USA) and NINS (Japan), 
together with NRC (Canada), MOST and ASIAA (Taiwan), and KASI (Republic of Korea), in 
cooperation with the Republic of Chile. The Joint ALMA Observatory is operated by ESO, AUI/
NRAO and NAOJ.
This research has
made use of NASA's Astrophysics Data System Bibliographic Services and the
SIMBAD database, operated at CDS, Strasbourg, France.
      
\end{acknowledgements}

\bibliographystyle{aa}
\bibliography{i04166_alma_c8.bib}

\onecolumn
\begin{appendix}
\section{Additional images}

\begin{figure}[!h]
       \centering
    \includegraphics[width=0.7\textwidth]{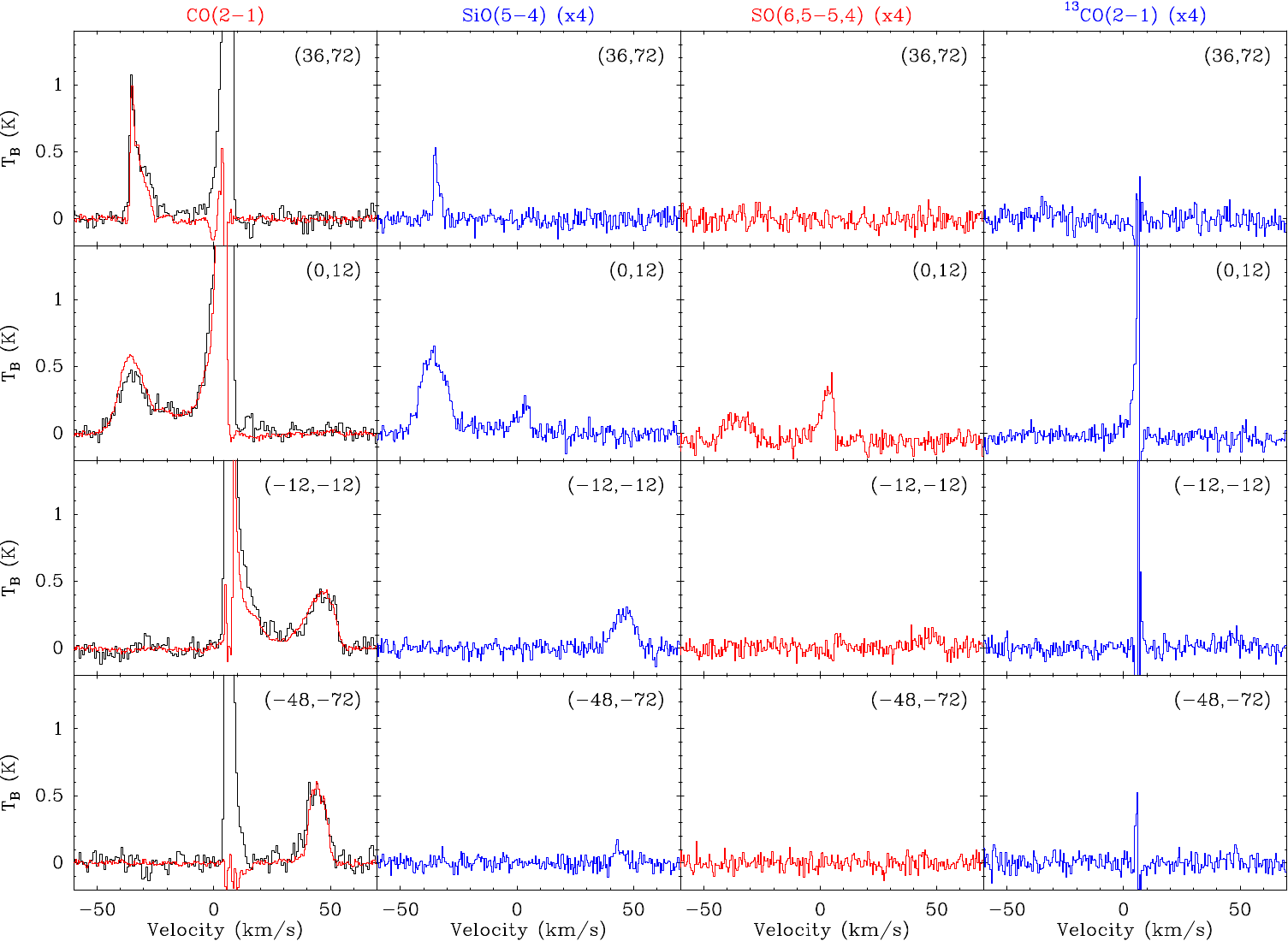}
        \caption{Spectra of the four transitions observed
        with ALMA convolved to an angular resolution of $11''$ 
        to simulate an observation with 
        the IRAM 30m telescope. For CO(2--1), the black lines represent
        actual IRAM 30m observations from \cite{taf04}. We note how
        the ALMA data recover all the CO(2--1) single-dish flux
        in the EHV regime ($|V-V_0|>30$~km~s$^{-1}$, where 
        $V_0 = 6.7$~km~s$^{-1}$ is the cloud systemic velocity),
        but lose some flux within 10~km~s$^{-1}$ of the ambient cloud.
        The offsets given in the top right corners are in 
        arcseconds with respect to the position of IRAS 04166.}
        \label{fig_convol}
\end{figure}

\begin{figure}[h]
        \centering
    \includegraphics[width=0.7\hsize]{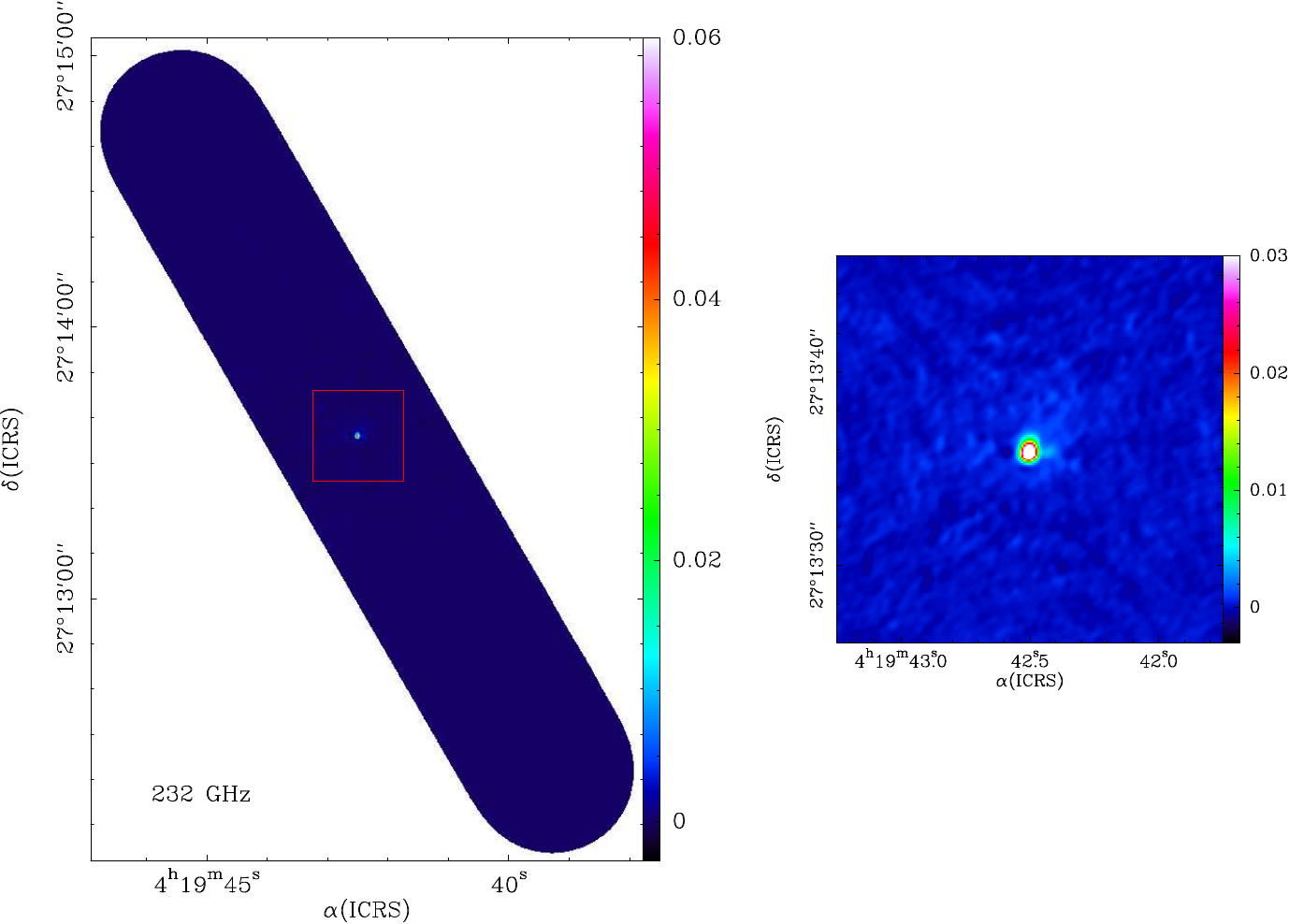}
        \caption{Maps of the 232~GHz continuum emission toward the IRAS 04166 outflow. {\em Left:} Full view of the ALMA 
        mosaic showing a single point-like source coincident with the IRAS position. {\em Right: } Expanded
        view of the central $20''\times 20''$ (red box in the left panel) with a stretched scale to emphasize
        the presence of sidelobes indicative of missing extended emission surrounding the unresolved central source.
        All intensities are in Jy~beam$^{-1}$.}
        \label{fig_cont}
\end{figure}

\begin{figure*}
        \centering
    \includegraphics[width=\hsize]{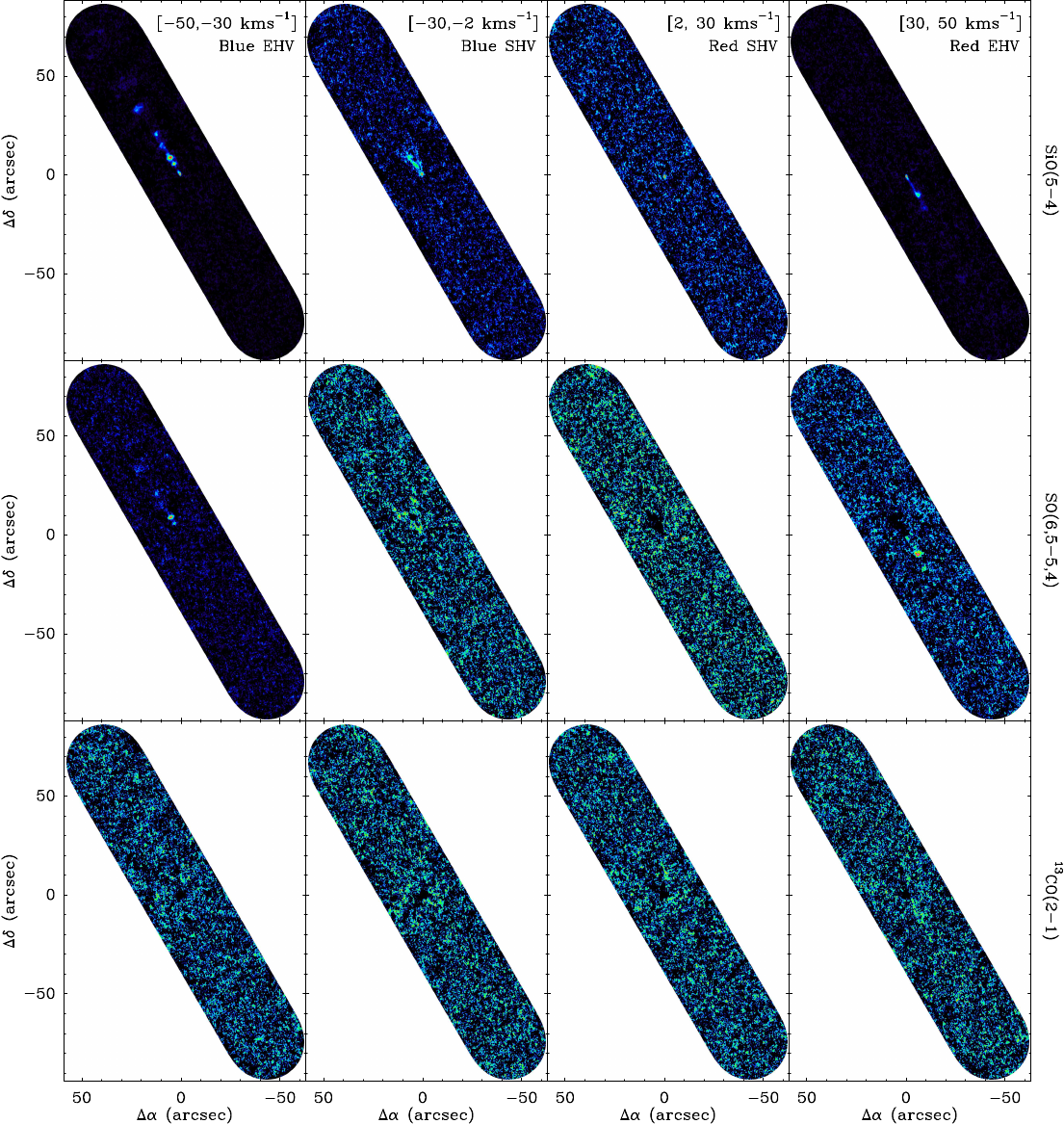}
        \caption{Maps of integrated intensity for the additional transitions observed
        with ALMA integrated over the EHV and SHV regimes. 
        The coordinates and velocity ranges are as in Fig.~\ref{fig_flow_4v}.
        For each panel, the intensity scale is linear with limits adjusted to show
        maximum contrast. We note the weaker signal compared to the CO(2--1) emission.}
        \label{fig_flow_others}
\end{figure*}

\twocolumn

\begin{figure*}
    \sidecaption
    \includegraphics[width=12cm]{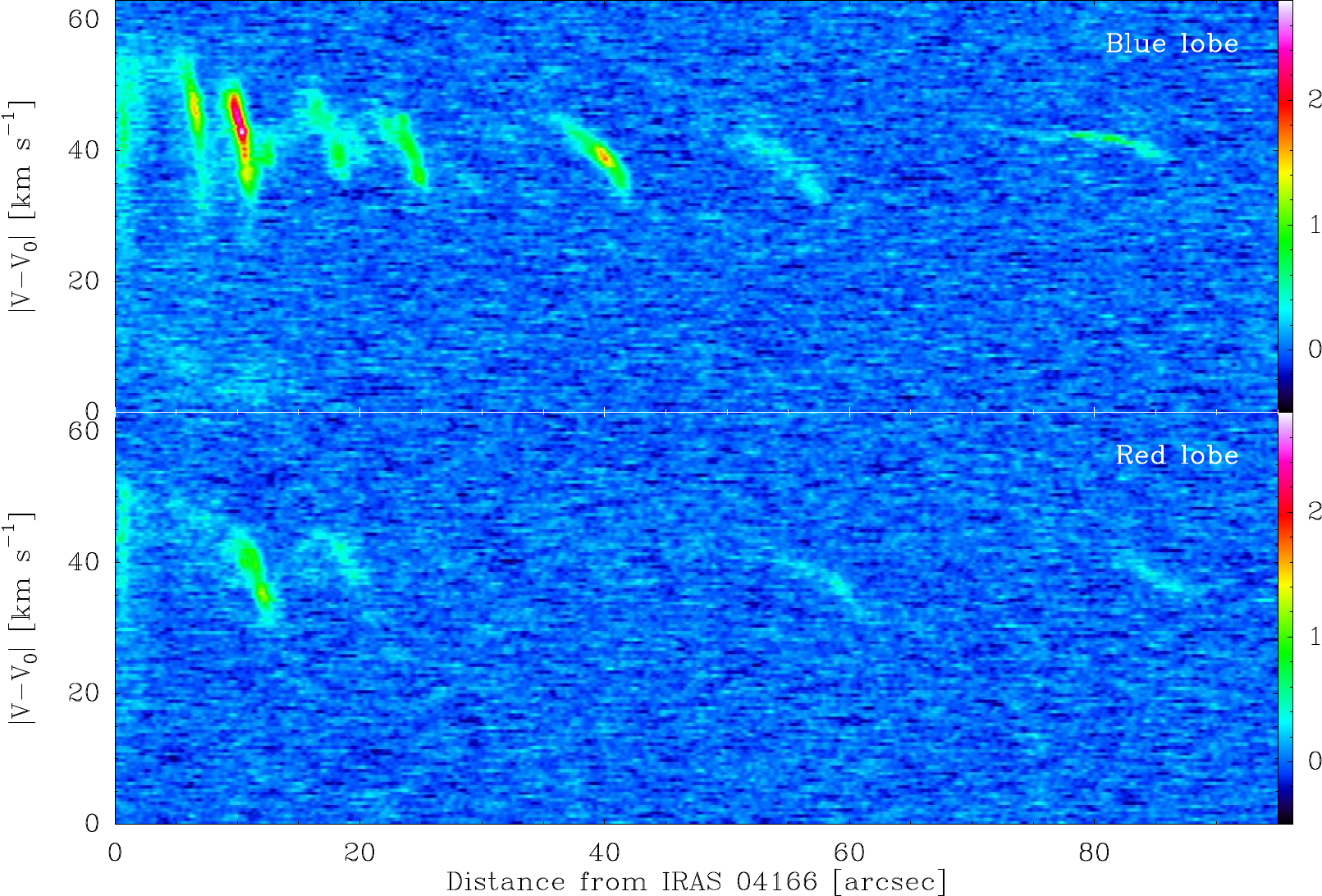}
    \caption{Position--velocity diagrams of the
SiO(5--4) emission along the axis of the blue
(top) and red (bottom) lobes of the IRAS 04166
outflow. The labels and scales are as in Fig.~\ref{fig_pv_along}}
        \label{fig_pv_sio5}
\end{figure*}

\begin{figure*}
\centering
   \includegraphics[width=0.9\hsize]{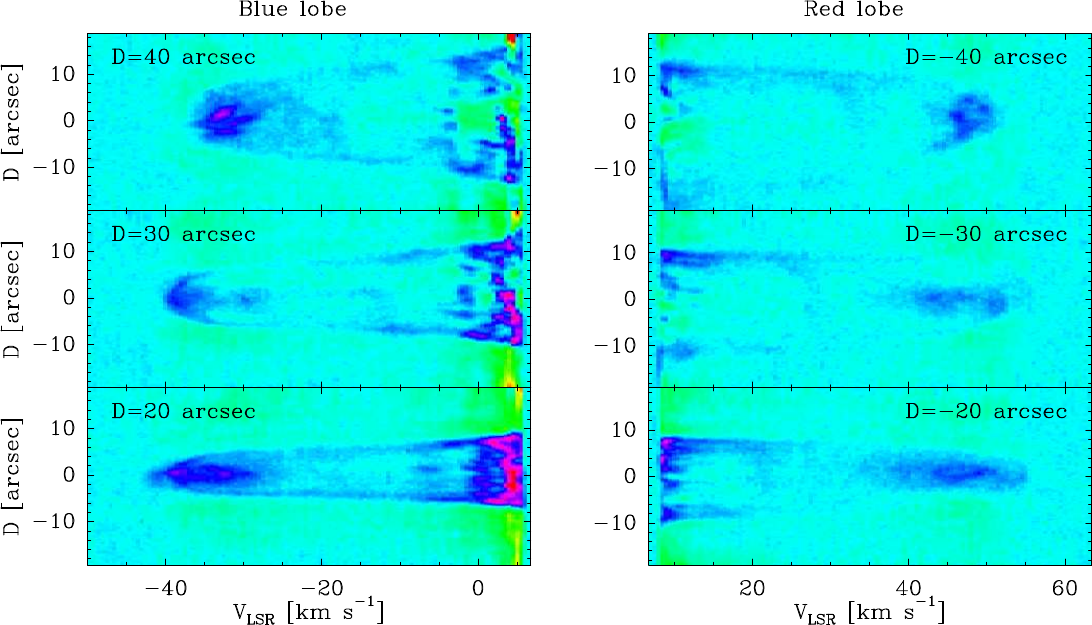}
        \caption{Position--velocity diagrams of the
CO(2--1) emission perpendicular to the blue
and red lobes of the IRAS 04166 outflow
at $20''$, $30''$, and $40''$ from the protostar. 
The weak emission inside the lobes 
hints at the presence of elliptical cavities
indicative of nested outflow shells.}
        \label{fig_pv_perp}
\end{figure*}

\begin{figure*}
  \sidecaption
   \includegraphics[width=0.55\hsize]{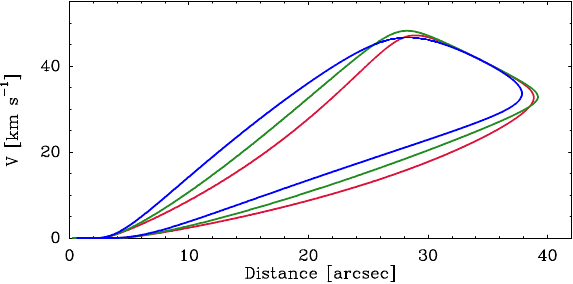}
        \caption{Position--velocity diagrams for single-event ejection models
        using three different density laws: $r^{-2}$ (blue), $r^{-1}$ (green),
        and no radial gradient (red). For easier comparison, all models 
        have the same density at $r=4''$.
        We note the similar shape of the 
        diagrams and their overlap toward the top right region
        corresponding to the EHV component. 
        This similarity indicates that the choice of density gradient
        has a minimal effect in the determination of the ejection model 
        presented in Sect.~\ref{sect_model_results}. 
        We also note  that the steeper the gradient, the faster the gas moves along the 
        PV fingers. This is a consequence of the lower density of the flow material
        encountered by the ejection.
        All models assume an inclination angle of $47^\circ$ 
        with respect to the line of sight, which corresponds to our 
        preferred choice for the IRAS 04166 outflow (see text).
        }
        \label{fig_dens_test}
\end{figure*}

\end{appendix}

\end{CJK*}
\end{document}